\documentclass[AMA,STIX1COL]{WileyNJD-v2}
\usepackage{graphicx}
\usepackage{epstopdf}
\articletype{Article Type}%

\received{26 April 2016}
\revised{6 June 2016}
\accepted{6 June 2016}

\begin{document}

\title{Assessment of intra-tumor heterogeneity in a two-dimensional vascular tumor growth model}

\author{Ioannis Lampropoulos}

\author{Michail Kavousanakis*}

\authormark{Lampropoulos and Kavousanakis}

\address{\orgdiv{School of Chemical Engineering}, \orgname{National Technical University of Athens}, \orgaddress{\state{Iroon Polytechneiou 9, Zografou 157 80, Athens}, \country{Greece}}}

\corres{*Michail E. Kavousanakis, School of Chemical Engineering, National Technical University of Athens, Iroon Polytechneiou 9, Zografou 15780, Athens, Greece. \email{mihkavus@chemeng.ntua.gr}}

\abstract[Summary]{
We present a two-dimensional continuum model of tumor growth, which treats the tissue as a composition of six distinct fluid phases; their dynamics are governed by the equations of mass and momentum conservation.
Our model divides the cancer cells phase into two sub-phases depending on their maturity state.
The same approach is also applied for the vasculature phase, which is divided into young sprouts (products of angiogenesis), and fully formed-mature vessels.
The remaining two phases correspond to healthy cells and extracellular material (ECM).
Furthermore, the model foresees the existence of nutrient chemical species, which are transferred within the tissue through diffusion or supplied by the vasculature (blood vessels).
The model is numerically solved with the Finite Elements Method and computations are performed with the commercial software Comsol Multiphysics \textsuperscript \textregistered.
The numerical simulations predict that mature cancer cells are well separated from young cancer cells, which form a protective shield for the growing tumor.
We study the effect of different mitosis and death rates for mature and young cancer cells on the tumor growth rate, and predict accelerated rates when the mitosis rate of young cancer cells is higher compared to mature cancer cells.
}


\maketitle

\footnotetext{\textbf{Abbreviations:} ITH, Intra-tumor heterogeneity; ECM, extracellular material; FEM, finite elements method}

\section{Introduction}\label{intro}
Malignant tumors are dynamic, self-organizing, and heterogeneous biological structures.
During their course, tumors become more heterogeneous and composed of a diverse collection of cells with different phenotypic behaviors, whether they belong in the same, or different tumors \cite{Fisher:2013}.
Unfortunately, evidence suggests that heterogeneity provides fuel for resistance to various therapies, and it is often the cause for treatment failure and cancer relapse due to the selective advantage it potentially provides \cite{Marusyk:2010,Marusyk:2012}.
Tumor heterogeneity is mainly classified into two major categories: (a) Inter-Tumor Heterogeneity, which refers to the divergent behavior
between different tumors of the same type and (b) Intra-Tumor Heterogeneity, which indicates differences in cancer cell
sub-populations within the same tumor, as a result of an ongoing evolutionary process accelerated by the piled up genetic and epigenetic mutations \cite{Fisher:2013,Loeb:2011,buck:2017}.
A third heterogeneity classification, called Inter-Site Heterogeneity, refers to the formation of heterogeneous tumors in the body of the same patient \cite{Piraino:2019}.
In this study, the focus is steered towards Intra-Tumor Heterogeneity (ITH), i.e. the observation of different sub-populations within the same malignant neoplasm.

ITH is a widely observed phenomenon in cancer research.
All kinds of malignant tumors
consist of cells that are inherently unstable, prone to mutations and highly diverse.
ITH is caused by genetic and epigenetic mutations, corresponding to differences in the phenotype of the cells constituting the same tumor \cite{Losic:2020,Gay:2016,Mcgranahan:2017}.
Indeed, there is evidence suggesting that ITH dynamics are not entirely attributed to gene mutations \cite{Losic:2020,Jamal:2017}.
Regardless of the causes, ITH has a radically negative impact on the success rate of cancer treatment, being a major contributor to a tumor’s resistance to targeted therapy \cite{Gay:2016}.
Cells belonging to different sub-populations antagonize each other and the healthy cells,
in a struggle to secure precious nutrients for their needs.
Because of that, tumors evolve in a fashion that could be considered Darwinian and cellular heterogeneity is a decisive factor
for the survival of such an unstable ecosystem \cite{Nowell:1976,greaves:2012}.
Heterogeneity increases the probability for resistance to therapy and is often associated to various factors of the tumor’s
immediate microenvironment.
Hinohara and Polyak \cite{Hinohara:2019} have studied both genetic and transcriptomic heterogeneity and discovered that drug tolerance in cancer
cells can be characterized by specific and reversible epigenetic states.
In this context, they have shown that targeting heterogeneity can increase treatment efficacy, while
fighting the rise of new and resilient cell populations.
Cancer heterogeneity has increasingly attracted the scientific community’s attention during the last few years,
as shown by the continuously increasing number of relative theoretical and experimental studies.
Baslan et al. \cite{Baslan:2020} detected cancer cells of the same tumor carrying different numbers of copies of the same genetic material.
Losic et al. \cite{Losic:2020} studied ITH in liver cancer, concluding that ITH has critical impact in cancer treatment, since sampling a
particular sub-population from a patient’s tumor can lead to false assumptions about the state of the illness and derail treatment \cite{Losic:2020,Jamal:2017}.
McDonald et al. \cite{Mcdonald:2019} concluded that ITH correlates with a lesser immune response and an overall smaller survival rate in breast cancer.
Gay et al. \cite{Gay:2016} suggest that cancer heterogeneity is heavily influencing cancer treatment, as it is inevitable that tumors will show heterogeneous sub-populations of cells with different responses to treatments.
Unless this is taken into account, the resistant sub-populations may survive target treatment and repopulate, breeding a new tumor more resistant compared to its predecessor.
It goes without saying that, quantifying ITH is a big opportunity to improve the success rate of cancer treatments.
ITH is also linked with the tendency of cells to metastasize.
Miller et al. \cite{Miller:1983} injected cells from the same mouse mammary tumor into different mice, in an effort to distinguish
variance in the injected tumors’ behavior.
Indeed, their research showed heterogeneous stability levels in the metastatic properties of the inserted cancerous sub-populations.
For further insights into the concept of tumor heterogeneity, the reader is referred to the reviews of Fisher et al \cite{Fisher:2013},
Gay et al. \cite{Gay:2016}, Hinohara and Polyak \cite{Hinohara:2019}, Zeng and Dai \cite{zeng:2019}, Koltai \cite{koltai:2017}, and Meacham and Morrison \cite{meacham:2013}.
In order to enhance our quantitative understanding of the clinical and experimental settings, as well as to illuminate the underlying bio-medical phenomena, several mathematical models have been developed for the simulation of heterogeneous tumor growth.
Zhao et al. \cite{Zhao:2014} developed an optimization algorithm in an effort to estimate the effect of various drug combinations on heterogeneous tumors.
Their predictions were based on the already measured performance of the drugs being administered one at a time on various cell sub-populations.
The efficacy of a drugs combination was calculated as a weighted average of its
single-drug counterparts on a homogeneous tumor.
Alvarez et al. \cite{Alvarez:2019} studied the tumor-immune dynamics of heterogeneous tumors through a non-linear mathematical model of cancer
immuno-surveillance.
The model takes into account ITH by distinguishing two cancer cell sub-populations based on their immunogenicity, i.e. their
ability to trigger an immune response.
On the topic of immunosurveillance one can also refer to the reviews of Dunn et al. \cite{Dunn:2002,Dunn:2004}.
Harris et al. \cite{Harris:2019} contemplated upon the merits of different approaches of heterogeneous tumor modeling, namely agent-based models,
population dynamics and multiscale models.
Stamatelos et al. \cite{Stamatelos:2019} presented a 3D hybrid image-based modeling framework and demonstrated the correlation between the underlying
irregularities in vascular growth and the emergence of ITH.
Heterogeneity is also intertwined with stemness.
In a tumor where ITH is expressed, there have been confirmed cases of cellular sub-populations expressing a phenotype
resembling that of stem cells \cite{Lapidot:1994,Al:2003,OBrien:2007,Ricci:2007,Singh:2004,Collins:2005}.
Tripathi et al. \cite{Tripathi:2020} developed a model targeting epithelial-mesenchymal heterogeneity, a phenomenon observed amongst
the cells produced through mitotic division of a cancer cell.
Their computations succeeded in capturing the changes observed in the volume fractions of different cancerous phenotypes
coexisting in tumors of the prostate.
Furthermore they were able to predict the tumor’s composition of different phenotypes in cells sub-populations.
Interestingly enough, they concluded that taking simultaneous action on both epithelial and mesenchymal cells will possibly increase
the likelihood for tumor growth restriction.
In this study, the heterogeneity element we focus on is the distinction of cancer cells based on their maturity state.
As it is going to be further explained in later paragraphs, the cells that are presented in the model are divided into
two sub-populations: the young and mature cells.
The state of maturity in cancerous populations is an emerging topic in the medical science.
Differentiation therapy is an aspiring method of treating cases of aggressive tumors by inducing cancer
cells maturation and differentiation through the insertion of the appropriate chemical agents into the tumor’s microenvironment \cite{Wang:2013,Leszczyniecka:2001}.
Of course, maturity in cancer cells has not been studied only in relation with the above technique.
Sontag et al. \cite{Sontag:2020} studied the possible correlation between a cancer cell’s thermal resistance and its maturity state,
posing the initial hypothesis that cancer stem cells found in human hepatocellular carcinoma are exhibiting higher resilience
to high temperatures, which are produced in hyperthermic techniques often used to treat hepatocellular carcinoma \cite{Sontag:2020,Naeem:2018},
than that of the non-stem and mature cells.
It is beyond doubt that vasculature also holds a very important role in the phenomenon of cancer growth.
Its presence is essential in malignant tumors as vessels generated through angiogenesis are the prelude to metastasis \cite{Pries:2009}\cite{Folkman:1971}.
Heterogeneity is also encountered in the vasculature generated during
angiogenesis \cite{Alvarez:2019,Bergers:2003,Folkman:1989}.
Heterogeneity in the surrounding vasculature of the tumor can be considered a leading cause for ITH and the irregularities
it can cause may hinder drug delivery in the tumor area \cite{Stamatelos:2019,junttila:2013}.
Vasculature produced internally by the cancer cells differ significantly from usual vasculature \cite{Pries:2009,Jain:2001,gazit:1995,baish:1996}.
Pries et al. \cite{Pries:2009} performed a series of simulations in an effort to correlate vessels structural properties to oxygen distribution in
the tumor area and response to different types of stimuli.
Vessels produced through angiogenesis are immature and permeable, among other characteristics \cite{Pries:2009,dewhirst:2007,sorg:2005}.
In the present study, we attempt to incorporate the different stages of maturity by dividing the vascular phase into two sub-phases: the young and mature vessels.
As it will be presented in the next paragraphs, these two phases exhibit different properties in an attempt to simulate the experimental
observations about microvasculature and hypoxic regions within the tumor.
The model presented in this paper is an extension of the model of Hubbard and Byrne \cite{Hubbard:2013}, which simulates the growth of a vascular tumor and incorporates interactions between cancer cells, healthy cells, vasculature,
extracellular material and the supplied nutrient.
Here, we incorporate ITH by modeling the existence of distinct sub-populations of cancer cells, as well as of heterogeneous vasculature, and quantify their effect on the overall dynamics of a growing tumor.
We also study the spatial tumor heterogeneity, by inspecting the relative position of different sub-populations within the tissue, which hosts healthy, cancer cells, vessels and extracellular matrix material (ECM).
The model is numerically solved with the method of Finite Elements, and computations are performed with Comsol Multiphysics \textsuperscript \textregistered.

The present paper is structured as follows: in Section~\ref{modeldev}, we present the details of the developed computational model, which incorporates cancer cell, and vessel heterogeneity during the growth of a vascular tumor.
In Section~\ref{results}, we present the results of simulations for different scenarios, where cancer cell heterogeneity is studied under the prism of different mitosis, and death rates, as well as for different material properties of the modelled phases.
The final Section~\ref{discussion} is devoted to the discussion of the produced results and the suggestions for future research.

\section{Model development}\label{modeldev}
In this section, we describe a multiphase model of vascular tumor growth, which incorporates cancer cell, and vasculature heterogeneity.
It is a two-dimensional, continuous model treating the tissue as a composition of six distinct and interacting phases.
For each of the phases, we model the evolution of their volume fraction, and in particular the volume fraction of: (i) healthy cells, which is denoted with $\theta_h$, (ii) young cancer cells, denoted with $\theta_{yc}$, (iii) mature cancer cells, $\theta_{mc}$, (iv) young vessels, $\theta_{yv}$, (v) mature vessels, $\theta_{mv}$ and (vi) extracellular material (ECM), $\theta_{ecm}$.
The aforementioned phases are modelled as viscous fluids exhibiting macroscopic flow, and we denote their velocities and pressures with $\vec{u}_i=\left(u_i,v_i \right)$ and $p_i$, respectively for $i=h,yc,mc,yv,mv,ecm$.
The model also stipulates the presence of a diffusible nutrient, $c$, which is supplied by vessels (young and mature), is transported through diffusion within the tissue and gets consumed by normal and cancer cells for survival and proliferation purposes.
Volume fractions of all six phases, as well as their velocities and pressures are derived by applying to each phase the principle of mass and momentum balance with appropriate constitutive equations;
the concentration of the diffusible nutrient is computed by solving a reaction-diffusion equation.

\subsection{Mass balance equations}
Under the assumption that density in a living tissue is uniform and constant, the mass balances for the six phases are simplified as follows:

\begin{equation}\label{eq:genmassbalance}
  \frac{\partial \theta_i}{\partial t} + \nabla \cdot \left( \vec{u}_i \theta_i \right) = q_i, \mbox{ for } i=h,yc,mc,yv,mv,ecm.
\end{equation}

\noindent The term, $\nabla \cdot \left( \vec{u}_i \theta_i \right)$  formulates mass transfer through convection, $\theta_i$ is the volume fraction of phase $i$ and $\vec{u}_i$ its velocity.
The source term, $q_i$, is associated with processes including cell mitosis (proliferation), death, angiogenesis, vessel occlusion (in general processes triggering transfer of mass between phases).

\subsubsection{Healthy cells}
The mass balance for the healthy cells is formulated as follows:

\begin{equation}\label{eq:healthymb}
  \frac{\partial \theta_h}{\partial t} + \nabla \cdot \left( \vec{u}_h \theta_h \right) =
  k_{m,h} \theta_h \theta_{ecm} \left( \frac{c}{c_p+c} \right) - k_{d,h} \theta_h \left( \frac{c_{c_1} + c}{c_{c_2} + c} \right) ,
\end{equation}

\noindent where $k_{m,h}$ is the healthy cells mitosis rate constant, $k_{d,h}$ is the healthy cells death rate constant.
The first term on the right hand side of Equation~(\ref{eq:healthymb}) models mitosis, with $c_p$ denoting the nutrient concentration at which the mitosis rate becomes half maximal.
The second term on the right hand side of Equation~(\ref{eq:healthymb}) represents the death rate of healthy cells;
$c_{c_1},c_{c_2}$ denote threshold nutrient concentration values regulating the cellular death rate.
Finally, in order to ensure that the death rate increases as the nutrient concentration decreases, we set $c_{c_1}>c_{c_2}$ .

\subsubsection{Cancer cells}
The cancer cells are grouped into two sub-populations depending on their state of maturity (age).
The first age group includes young cancer cells and the second group includes mature cells, which originate from young cells through a maturation process.
In our model, mitosis and death rates of cancer cells can be altered depending on their age.
The mass balance equations for young and mature cancer cells are:

\begin{equation}\label{eq:youngtumormb}
  \frac{\partial \theta_{yc}}{\partial t} + \nabla \cdot \left( \vec{u}_{yc} \theta_{yc} \right) =
  \left( k_{m,yc} \theta_{yc} +k_{m,mc} \theta_{mc} \right) \theta_{ecm} \left( \frac{c}{c_p+c} \right) - k_{d,yc} \theta_{yc} \left( \frac{c_{c_1} + c}{c_{c_2} + c} \right) -k_{mat}^{cell} \theta_{yc},
\end{equation}

\begin{equation}\label{eq:maturetumormb}
  \frac{\partial \theta_{mc}}{\partial t} + \nabla \cdot \left( \vec{u}_{mc} \theta_{mc} \right) =
   - k_{d,mc} \theta_{mc} \left( \frac{c_{c_1} + c}{c_{c_2} + c} \right) +k_{mat}^{cell} \theta_{yc},
\end{equation}
\noindent where $k_{m,yc}$ and $k_{d,yc}$ denote the young cancer cells mitosis and death rate constant, respectively.
For mature cancer cells, the mitosis and death rate constants are denoted with, $k_{m,mc}$ and $k_{d,mc}$, respectively.
The first term on the right hand side of Equation~(\ref{eq:youngtumormb}) represents the mitosis rate, which depends on the nutrient concentration, $c$ and the volume fraction of ECM, $\theta_{ecm}$.
The mitosis rate is proportional to the volume fraction of cancer cells (both young and mature);
through the cell birth process young cancer cells will be produced.
For simplicity purposes, we set the threshold nutrient concentration values, $c_p, c_{c_1}, c_{c_2}$ for mitosis and death rates to be identical for normal and cancer cells.
In order to take into account that cancer cells proliferate more rapidly, and are less prone to death compared to healthy cells, we set $k_{m,yc}, k_{m,mc} > k_{m,h}$ and $k_{d,yc}, k_{d,mc} < k_{d,h}$.
Finally, we formulate the maturation process through which young cells become mature ones, with the last term of the right hand side of Equations~(\ref{eq:youngtumormb}) and (\ref{eq:maturetumormb});
maturation is formulated as a first order reaction proportional to the volume fraction of young cells, $\theta_{yc}$, and $k_{mat}^{cell}$ is the cell maturation constant rate.
We define the value of the cell maturation rate constant, $k_{mat}^{cell}$ through:

\begin{equation}\label{eq:propmatk12}
k_{mat}^{cell} = \beta \cdot k_{m,yc},
\end{equation}

\noindent where $\beta$ is a constant, and $0<\beta << 1$, i.e., the maturation rate constant is significantly smaller compared to the mitosis rate constant of young cancer cells.

\subsubsection{Vasculature}
An additional source of intrinsic heterogeneity is the differentiation of blood vessels.
We adopt a similar approach with the division of the cancerous phase into young and mature cells, and separate the blood vessels phase into two distinct phases: young vessels (sprouts formed due to angiogenesis) and mature vessels.
The mass balance equations for young and mature vessels respectively are:

\begin{equation}\label{eq:youngvesselsmb}
\begin{split}
  \frac{\partial \theta_{yv}}{\partial t} + \nabla \cdot \left( \vec{u}_{yv} \theta_{yv} \right) & =
  k_{ang} \left( \theta_h + \theta_{yc} + \theta_{mc} \right) \left( \theta_{yv} + \theta_{mv} \right) \left( \frac{\theta_{ecm}}{\epsilon + \theta_{ecm}} \right)  \frac{c}{\left(c_a + c \right)^2} \\
  & -k_{occ}^{yv} \theta_{yv} \mathcal{H} \left( \theta_h p_h + \theta_{yc} p_{yc} + \theta_{mc} p_{mc} -p_{crit}^{yv},h \right) -k_{mat}^{ves} \theta_{yv},
  \end{split}
\end{equation}

\begin{equation}\label{eq:maturevesselsmb}
  \frac{\partial \theta_{mv}}{\partial t} + \nabla \cdot \left( \vec{u}_{mv} \theta_{mv} \right) =
  -k_{occ}^{mv} \theta_{mv} \mathcal{H} \left( \theta_h p_h + \theta_{yc} p_{yc} + \theta_{mc} p_{mc} - p_{crit}^{mv}, h \right) + k_{mat}^{ves} \theta_{yv},
\end{equation}
\noindent The first term of the right hand side in Equation~(\ref{eq:youngvesselsmb}) corresponds to the angiogenesis process, during which young vessels are produced.
We denote with, $k_{ang}$, the angiogenesis rate constant
and the angiogenesis rate is modelled to be proportional to the total volume fraction of healthy, young and mature cancer cells, as well as proportional to the total volume fraction of young and mature vessels.
The vessel growth is also assumed to get promoted by the presence of ECM and the diffusible nutrient, and reaches a maximum rate for high values of ECM ($\theta_{ecm}$) and nutrient concentration values;
$\epsilon$ corresponds to the ECM volume fraction at which the angiogenesis rate becomes half maximal, and $c_a$ corresponds to the nutrient concentration at which the angiogenesis rate becomes maximal.
We note that the angiogenesis term appears only as a source term for young vessels.
Vessel occlusion is present in both young (second term of the right hand side in Equation~(\ref{eq:youngvesselsmb})) and mature vessels (first term of the right hand side in Equation~(\ref{eq:maturevesselsmb}).
For both phases the occlusion rate depends proportionally to the volume fraction of vessels (young and mature, respectively), and gets triggered when the pressure exerted by the surrounding phases ($\theta_h p_h + \theta_{yc} p_{yc} + \theta_{mc} p_{mc}$) exceeds a threshold value, denoted with $p_{crit}^{yv}$ for young and $p_{crit}^{mv}$ for mature vessels.
For this purpose, we model $\mathcal{H}$ to be a smooth transition function:
\begin{equation}\label{eq:stepfunction}
\mathcal{H} \left( \theta_h p_h + \theta_{yc} p_{yc} + \theta_{mc} p_{mc} -p_{crit}, h \right) = \frac{1}{2} \left[ 1+ \tanh \left( \frac{\theta_h p_h + \theta_{yc} p_{yc} + \theta_{mc} p_{mc}-p_{crit}}{h_{sm}} \right) \right],
\end{equation}

\noindent where $h_{sm}$ is a smoothness parameter, and $p_{crit}=p_{crit}^{yv}$ or $p_{crit}^{mv}$, for young and mature vessels, respectively.
The major differences in the functionality of vessels belonging to different age groups refer to their ability to withstand vascular occlusion, with mature vessels being more resilient compared to young sprouts.
Thus, we set different pressure threshold values, namely: $p_{crit}^{mv}$ for mature vessels, and $p_{crit}^{yv}$ for young vessels with $p_{crit}^{yv} < p_{crit}^{mv}$.
In addition, we consider that the rate constant of occlusion for mature vessels, $k_{occ}^{mv}$ is smaller compared to the occlusion rate constant of young vessels, $k_{occ}^{yv}$, i.e.: $k_{occ}^{mv} < k_{occ}^{yv}$
Finally, we model the maturation process of blood vessels as a first order reaction, which is proportional to the volume fraction of young vessels, and $k_{mat}^{ves}$ being the maturation rate constant.

\subsubsection{Extracellular Material}
ECM is a passive medium hosting material that is used for proliferation purposes, or produced from cellular death and vascular occlusion (necrotic material).
By assuming that the studied system is closed (no external replenishment of resources) the sum of source terms, $q_i$ for all phases must sum up to $0$:
\begin{equation}\label{eq:convsource}
  \sum_{i} q_i= 0, \mbox{ for } i=h,yc,mc,yv,mv,ecm.
\end{equation}
\noindent  In accordance to the assumption above, the source term for ECM is defined as:

\begin{equation}\label{eq:ecmmb}
  \frac{\partial \theta_{ecm}}{\partial t} + \nabla \cdot \left( \vec{u}_{ecm} \theta_{ecm} \right) =
  q_{ecm} = -\sum_{i} q_i, \textrm{ for } i=h,yc,mc,yv,mv.
\end{equation}

In practice, we adopt the assumption that the tissue has no voids:

\begin{equation}\label{eq:novoid}
  \sum_{i} \theta_i = 1, i=h,yc,mc,yv,mv,ecm,
\end{equation}
\noindent and calculate the volume fraction of ECM, through Equation~(\ref{eq:novoid}), instead of computing $\theta_{ecm}$ from Equation~(\ref{eq:ecmmb}).

\subsubsection{Diffusible nutrient}
We assume that the timescales of transport and reaction processes for nutrient are considerably shorter (minutes) compared to the timescales of processes associated with fluid phase changes (days or weeks);
under this assumption, the nutrient is in a quasi-steady state and the general form of its mass balance equation can be written as follows:

\begin{equation}\label{eq:nutrmb}
  -D_c \nabla^2c = q_c.
\end{equation}
\noindent $D_c$ is the diffusion coefficient for the nutrient and $q_c$ its source term equal to:

\begin{equation}\label{eq:nutrientsource}
  q_c = k_{rep} (\theta_{yv} + \theta_{mv})(c_v-c) - (k_{c,h} \theta_h c + k_{c,yc} \theta_{yc} c + k_{c,mc} \theta_{mc} c)- \left[ k_{cm,h} \theta_h  + k_{cm,yc} \theta_{yc} + k_{cm,mc} \theta_{mc}   \right] \theta_{ecm} \left( \frac{c}{c_p+c} \right);
\end{equation}

\noindent $k_{rep}$ is the nutrient replenishment rate constant for the replenishment performed by the local vasculature, $c_v$ is the nutrient concentration inside said vasculature  $k_{c,i},i=h,yc,mc$, denote the consumption rate constant for cell sustenance of healthy and cancer cells, respectively and $k_{cm,i},i=h,yc,mc$, the rate constants for the consumption of nutrient for the purposes of mitosis for healthy and cancer cells, respectively.
The value of $k_{cm,i}$ is given by the formula:

\begin{equation}\label{eq:k7i}
  k_{cm,i} = \frac{k_{cm,h} \cdot k_{m,i}}{k_{m,h}}, i=h,yc,mc.
\end{equation}
Again, we stress that the nutrient contribution to the overall volume of the tissue is  neglected and that its mass transport is induced by diffusion.

\subsection{Momentum balance equations}
Assuming creeping flow, we write the general form of momentum balance equation for each phase,$i$:

\begin{equation}\label{eq:mombalgeneral}
  \nabla \cdot \left( \theta_i \boldsymbol{\sigma}_i \right) + \vec{F}_i = \vec{0}, \mbox{ for } i=h,yc,mc,yv,mv,ecm,
\end{equation}

\noindent where $\boldsymbol{\sigma}_i$ denotes the stress tensor of phase, $i$, and $\vec{F}_i$ denotes the momentum source terms, which include the effects of pressure and inter-phase drag:

\begin{equation}\label{eq:forcegeneral}
  \vec{F}_i = p_i \boldsymbol{I} \nabla \theta_i + \sum_{j,j\neq i} d_{i,j} \theta_i \theta_j \left( \vec{u}_j - \vec{u}_i \right), \mbox{ for } i=h,yc,mc,yv,mv,ecm.
\end{equation}

Considering each fluid phase as a viscous and compressible fluid, the stress tensor is:

\begin{equation}\label{eq:sigmatensor}
  \boldsymbol{\sigma}_i = -p_i \boldsymbol{I} + \mu_i \left( \nabla \vec{u}_i + \left( \nabla \vec{u}_i \right)^T \right) -\frac{2}{3} \mu_i \left( \nabla \cdot \vec{u}_i \right) \boldsymbol{I}.
\end{equation}

In Equations~(\ref{eq:forcegeneral})-(\ref{eq:sigmatensor}), $p_i$ is the phase pressure, $d_{i,j}$ is the drag coefficient due to the relative movement of the phases, $i$ and $j$, surfaces; $\mu_i$ is the dynamic viscosity of phase $i$.
For the calculation of both pressures and velocity fields for each phase, a continuity equation for the phase mixture is required,
and can be produced by the summation of the mass balance equations of all phases (Equation (\ref{eq:genmassbalance}) for $i=h,yc,mc,yv,mv,ecm$):

\begin{equation}\label{eq:conservmom}
  \sum_{i} \nabla \cdot \left( \vec{u}_i \theta_i \right) = 0, i=h,yc,mc,yv,mv,ecm.
\end{equation}
\noindent Furthermore, appropriately defined constitutive relations correlating the different phase pressures are also required for the closure of the system of equations.
In particular, we set the vascular phase pressures, $p_{yv}, p_{mv}$, to be equal to a reference pressure, $p_{ref}=0$, (externally imposed pressure in the vasculature, which is set to 0), and the rest of the pressures are related through:

\begin{equation}\label{eq:eos}
  p_h = p_{yc} = p_{mc} = p_{ecm} + \Sigma \left( \theta_h + \theta_{yc} + \theta_{mc} \right).
\end{equation}

\noindent $\Sigma(\theta)$ is a function calculating the increase in pressure exerted by cells whenever their local density exceeds their natural value $\theta^*$ (the cellular volume fraction value in a healthy tissue):

\begin{equation}\label{eq:Sigma}
  \Sigma(\theta) = \begin{cases}
                      \frac{\Lambda (\theta-\theta^*)}{\left(1-\theta\right)^2}, & \mbox{if } \theta \geq \theta^*\\
                      0, & \mbox{otherwise}.
                    \end{cases}
\end{equation}

\noindent where $\Lambda$ is a tension constant measuring the tendency of cells to restore their natural density.

\subsection{Non-dimensionalised model}

Before proceeding to the presentation of the results of the model, we convert the variables to a dimensionless form.
Volume fractions $\theta_i$ are already dimensionless so the rest of the variables are non-dimensionalised as follows:

\begin{equation}\label{eq:nondimensional}
  t^\prime = k_{m,h} t, \vec{x^\prime} = \frac{\vec{x}}{R_o}, \vec{u^\prime}_i = \frac{\vec{u}_i}{k_{m,h}R_o}, p^\prime_i = \frac{p_i}{ \Lambda}, c^\prime=\frac{c}{c_v},
\end{equation}

\noindent where $R_o$ is a typical length scale; here $R_o$ is the radius of the initial cancerous seed.
The resulting dimensionless mass balances read:
\begin{equation}\label{eq:massbalancedim_healthy}
  \frac{\partial \theta_h}{\partial t} + \nabla \cdot \left( \theta_h \vec{u}_{h}^\prime \right) = \theta_h \theta_{ecm} \left( \frac{c^\prime}{c_p^* + c^\prime} \right) -k_{d,h}^*\theta_h \left( \frac{c_{c_1}^* + c^\prime}{c_{c_2}^* + c^\prime} \right),
\end{equation}
\begin{equation}\label{eq:massbalancedim_yc}
  \frac{\partial \theta_{yc}}{\partial t} + \nabla \cdot \left( \theta_{yc} \vec{u}_{yc}^\prime \right) = (k_{m,yc}^* \theta_{yc} + k_{m,mc}^* \theta_{mc}) \theta_{ecm} \left( \frac{c^\prime}{c_p^*+c^\prime} \right) - k_{d,yc}^* \theta_{yc} \left( \frac{c_{c_1}^*+c^\prime}{c_{c_2}^*+c^\prime} \right) - k_{mat}^{cell^*} \theta_{yc},
\end{equation}
\begin{equation}\label{eq:massbalancedim_mc}
\frac{\partial \theta_{mc}}{\partial t} + \nabla \cdot \left( \theta_{mc} \vec{u}_{mc}^\prime \right) = -k_{d,mc}^* \theta_{mc} \left( \frac{c_{c_1}^*+c^\prime}{c_{c_2}^*+c^\prime} \right) + k_{mat}^{cell^*} \theta_{yc},
\end{equation}
\begin{equation}\label{eq:massbalancedim_yv}
  \begin{split}
  \frac{\partial \theta_{yv}}{\partial t} + \nabla \cdot \left( \theta_{yv} \vec{u}_{yv}^\prime \right) &= k_{ang}^* (\theta_h + \theta_{yc} + \theta_{mc}) (\theta_{yv} + \theta_{mv}) \left( \frac{\theta_{ecm}}{\epsilon + \theta_{ecm}} \right)  \frac{c^\prime}{(c_a^* + c^\prime)^2}  \\
  &- k_{occ}^{yv^*} \theta_{yv} \mathcal{H} ( \theta_h p_h^\prime + \theta_{yc} p_{yc}^\prime + \theta_{mc} p_{mc}^\prime - p_{crit}^{yv^*}, h_{sm}^* ) - k_{mat}^{ves^*} \theta_{yv},
  \end{split}
\end{equation}
\begin{equation}\label{eq:massbalancedim_mv}
  \frac{\partial \theta_{mv}}{\partial t} + \nabla \cdot \left( \theta_{mv} \vec{u}_{mv}^\prime \right) = -k_{occ}^{mv^*} \theta_{mv} \mathcal{H} (\theta_h p_h^\prime + \theta_{yc} p_{yc}^\prime +\theta_{mc} p_{mc}^\prime - p_{crit}^{mv^*}, h_{sm}^* ) + k_{mat}^{ves^*} \theta_{yv},
\end{equation}

\begin{eqnarray}\label{eq:massbalancedim_ecm}
  \theta_{ecm} = 1-\theta_h-\theta_{yc}-\theta_{mc} -\theta_{yv} - \theta_{mv},
\end{eqnarray}

\noindent where:
\begin{equation*}
  k_{m,yc}^*=\frac{k_{m,yc}}{k_{m,h}}, k_{m,mc}^*=\frac{k_{m,mc}}{k_{m,h}},
\end{equation*}
\begin{equation*}
  k_{d,h}^* = \frac{k_{d,h}}{k_{m,h}}, k_{d,yc}^* = \frac{k_{d,yc}}{k_{m,h}}, k_{d,mc}^*=\frac{k_{d,mc}}{k_{m,h}},
\end{equation*}
\begin{equation*}
  k_{occ}^{mv^*} = \frac{k_{occ}^{mv}}{k_{m,h}}, k_{occ}^{yv^*}=\frac{k_{occ}^{yv}}{k_{m,h}}, k_{ang}^*=\frac{k_{ang}}{k_{m,h}c_v},
\end{equation*}
\begin{equation*}
  k_{mat}^{cell^*}=\frac{k_{mat}^{cell}}{k_{m,h}}, k_{mat}^{ves^*}=\frac{k_{mat}^{ves}}{k_{m,h}},
\end{equation*}
\begin{equation*}
  c_p^*=\frac{c_p}{c_v}, c_{c_1}^* = \frac{c_{c_1}}{c_v}, c_{c_2}^* = \frac{c_{c_2}}{c_v}, c_a^* = \frac{c_a}{c_v},
\end{equation*}
\begin{equation*}
  h_{sm}^* = \frac{h_{sm}}{\Lambda}, p_{crit}^{yv^*}=\frac{p_{crit}^{yv}}{\Lambda}, p_{crit}^{mv^*}=\frac{p_{crit}^{mv}}{\Lambda}.
\end{equation*}

Since the mass balance equations are hyperbolic, boundary conditions are only defined on segments of the domain's, $\Omega$ boundaries (denoted with $\partial \Omega$) where: $\vec{u}_i \cdot \vec{n} < 0$ (inflow segments), with
$\vec{n}$ denoting the outward-pointing unit normal vector of the boundary.
For boundaries with $\vec{u}_i \cdot \vec{n} < 0$, the imposed boundary conditions are:

\begin{equation}\label{eq:bctheta}
  \theta_i = \theta_i^{\infty}, i=h,yc,mc,yv,mv.
\end{equation}
The dimensionless mass balance equation for the nutrient is:

\begin{equation}\label{eq:nutrientdim}
\begin{split}
  -D_c^* \nabla^2 c^\prime &= (\theta_{mv} + \theta_{yv})(1-c^\prime)-(k_{c,h}^* \theta_h c^\prime + k_{c,yc}^* \theta_{yc} c^\prime + k_{c,mc}^* \theta_{mc} c^\prime) \\
  &- \left[ k_{cm,h}^* \theta_h  + k_{cm,yc}^* \theta_{yc} + k_{cm,mc}^* \theta_{mc}  \right] \theta_{ecm} \left( \frac{c^\prime}{c_p^* + c^\prime} \right),
  \end{split}
\end{equation}
\noindent where: $D_c^* = \frac{D_c}{k_{rep} R_o^2}, k_{c,h}^*=\frac{k_{c,h}}{k_{rep}}, k_{c,yc}^*=\frac{k_{c,yc}}{k_{rep}}, k_{c,mc}^*=\frac{k_{c,mc}}{k_{rep}}, k_{cm,h}^*=\frac{k_{cm,h}}{k_{rep} c_v}, k_{cm,yc}^*=\frac{k_{cm,yc}}{k_{rep} c_v}, k_{cm,mc}^*=\frac{k_{cm,mc}}{k_{rep} c_v}$.
The mass balance of nutrient is an elliptic partial differential equation, thus there is no need to identify inflow segments of $\partial \Omega$.
In this work, we impose Neumann type boundary condition for the nutrient concentration:

\begin{equation}\label{eq:nutrientbc}
  \nabla c^{\prime} \cdot \vec{n} = 0.
\end{equation}

The dimensionless form of momentum balances read:

\begin{equation}\label{eq:mombalancegendim}
  \sum_{j,j \neq i} d_{i,j}^* \theta_i \theta_j \left( \vec{u^\prime}_j - \vec{u^\prime}_i \right) -\theta_i \nabla \cdot \left( \Lambda^* p^\prime_i \boldsymbol{I} \right) + \nabla \cdot \left[ \theta_i \left[ \mu_i^* \left( \nabla \vec{u^\prime}_i +\left( \nabla \vec{u^\prime}_i \right)^T \right) -\frac{2}{3}\mu_i^* \left( \nabla \cdot \vec{u^\prime}_i \right) \boldsymbol{I} \right] \right]=\vec{0},
\end{equation}

\noindent for $i,j=h,yc,mc,yv,mv,ecm$.
Here $d_{i,j}^*=\frac{d_{i,j}}{d_{h,yc}}, \Lambda^* = \frac{\Lambda}{d_{h,yc}k_{m,h}R_o^2}, \mu_i^*=\frac{\mu_i}{d_{h,yc}R_o^2}$.
The boundary conditions for Equations~(\ref{eq:mombalancegendim}) are:

\begin{equation}\label{eq:bcmom}
  \boldsymbol{\sigma}_i \cdot \vec{n} = \vec{0}, \mbox{ for } i=h,yc,mc,yv,mv
\end{equation}

\noindent Since the specification of the normal stress along the whole boundary, $\partial \Omega$, is not possible for all phases,we specify the velocity for one phase (here the ECM) in order to obtain a unique solution:

\begin{equation}\label{eq:bcmom2}
  \vec{u}_{ecm} = \vec{0}.
\end{equation}

Finally, the dimensionless form of the continuity equation (Equation~(\ref{eq:conservmom})) reads:

\begin{equation}\label{eq:continuationdim}
  \sum_{i} \nabla \cdot \left( \theta_i \vec{u^\prime}_i \right) = 0, \mbox{ for } i=h,yc,mc,yv,mv,ecm,
\end{equation}

\noindent and the dimensionless equations of constitutive equations for pressures (Equation~(\ref{eq:eos}) are:

\begin{equation}\label{eq:eosdim}
  p^\prime_h = p^\prime_{yc} = p^\prime_{mc} = p^\prime_{ecm} + \Sigma'(\theta_h+\theta_{yc}+\theta_{mc}),
\end{equation}

\noindent with $\Sigma'(\theta)$:

\begin{equation}\label{eq:sigmadim}
\Sigma'(\theta) = \begin{cases}
                      \frac{ (\theta-\theta^*)}{\left(1-\theta\right)^2}, & \mbox{if } \theta \geq \theta^*\\
                      0, & \mbox{otherwise}.
                    \end{cases}
\end{equation}

\section{Results}\label{results}
All computations in the present study are performed in the environment of Comsol Multiphysics\textsuperscript \textregistered, which is based on the Finite Elements Method (FEM).
The tissue is modelled as circular domain with (dimensionless) radius, $R_{tissue}=16$.
The computational mesh is an unstructured mesh generated with the method of Delaunay Triangulation, and contains a total of approximately $20,000$ elements, which results in approximately $570,000$ degrees of freedom.
The required computational time on an Intel\textsuperscript \textregistered Core\textsuperscript {TM} i5-3360M CPU @ 2.80 GHz is approximately 11 hrs on average for each run presented in this section.

For simplification purposes, we drop the "$\prime$" and "*" symbols for all dimensionless variables.
All simulations are initialized by seeding a number of cancer cells on a healthy tissue, which can be considered being at an equilibrium state \cite{Hubbard:2013} and satisfies the following assumptions:

\begin{itemize}
  \item 	the fluid is a perfect mixture;
  as a result, the volume fraction of each phase is uniformly distributed and shows no macroscopic velocity ($\vec{u}_i=\vec{0},i=h,yv,mv,ecm$);
  \item 	the tissue shows no sign of contamination from cancer cells (young or mature) ($\theta_{yc}=\theta_{mc}=0$);
  \item 	healthy cells maintain a constant, natural cell density: in that state, no stress is exerted between cells, as this volume fraction value serves as the limit beyond which cell to cell interactions begin to occur ($\theta_h=\theta^*=0.6$);
  \item 	the pressure of each fluid phase is equal to zero (as a direct consequence of the above assumption)
\end{itemize}

Following the assumptions above, the system of Equations (\ref{eq:massbalancedim_healthy})-(\ref{eq:massbalancedim_ecm}) is reduced to the following set of nonlinear equations:

\begin{eqnarray}
  \theta^* \theta_{ecm} \frac{c}{c_p+c} -k_{d,h} \theta^* \frac{c_{c_1}+c}{c_{c_2}+c}&=& 0, \\
  -k_{occ}^{mv} \theta_{mc} \mathcal{H} \left( -p_{crit}^{mv}, h_{sm} \right) + k_{mat}^{ves} \theta_{yv} &=& 0, \\
  \theta^* + \theta_{mv} + \theta_{yv} + \theta_{ecm} &=& 1, \\
  k_{ang} \theta^* (\theta_{mv}+ \theta_{yv}) \frac{\theta_{ecm}}{\epsilon + \theta_{ecm}} \frac{c}{(c_a+c)^2} - k_{occ}^{yv} \theta_{yv} \mathcal{H} \left( -p_{crit}^{yv}, h_{sm} \right) - k_{mat}^{ves} \theta_{yv}&=& 0, \\
  (\theta_{mv} + \theta_{yv}) (1-c) - k_{c,h} \theta^* c - k_{cm,h} \theta^* \theta_{ecm} \frac{c}{c_p+c} &=& 0.
\end{eqnarray}

Solving the model for the volume fractions of $\theta_{mv}, \theta_{yv}, \theta_{ecm}$ and the non-dimensionalized concentration of $c$ yields the initial conditions of the studied system.
In order to obtain a unique solution (four unknown variables for a system of five equations), we also consider $k_{mat}^{ves}$  as an extra unknown variable.
Naturally, the results produced have to satisfy $\theta_{mv},\theta_{yv},\theta_{ecm},c \in (0,1)$ and $k_{mat}^{ves}>0$ in order to be considered as valid.

Using the parameter values presented in Table \ref{tab1}, the obtained solution is the following:

\begin{equation}\label{eq:initcond}
  \theta_{mv}(x,y,0)=0.015335, \theta_{ecm}(x,y,0)=0.38250,
\theta_{yv}(x,y,0)=0.0021625, c (x,y,0)=0.25320, k_{mat}^{ves}=0.029041.
\end{equation}

\subsection{Growing tumor with uniformly behaving sub-populations}
In this paragraph, we simulate the case of a growing tumor in which both young and mature cancer cells behave in an homogeneous manner, i.e. mature and young cancer cells share the same proliferation and death rate constants.
The initial population of cancer cells are seeded in a circular domain of radius, $R_o=1$ around the origin of the computational domain.
In particular, the initial cancerous seed is implanted according to the formula:

\begin{equation}\label{eq:initialseed}
  \theta_{yc}(x,y,0) = \begin{cases}
                      0.05 \cos^2 \left( \frac{\pi r}{2} \right), & \mbox{if } \sqrt{x^2+y^2} \leq R_o = 1\\
                      0, & \mbox{otherwise},
                    \end{cases}
\end{equation}

\noindent It is assumed that the initial cancerous seed consists exclusively of young cancer cells.
Consequently, the initial distribution of mature cancer cells is:

\begin{equation}\label{eq:initmaturetumor}
  \theta_{mc}(x,y,0)=0.
\end{equation}

The young cancer cells replace healthy cells, thus the initial condition for $\theta_h$ reads:

\begin{equation}\label{eq:inithealthy}
  \theta_h(x,y,0)=0.6 - \theta_{yc} (x,y,0).
\end{equation}

\noindent The initial condition of $\theta_{yv},\theta_{mv},\theta_{emc}$ and $c$ is provided from (\ref{eq:initcond}).
In this numerical experiment, the constants describing the two sub-populations of cancer cells are kept equal, so as to simulate a tumor with uniform growth and death rates.
The values of parameters are presented in Table \ref{tab1}.
For the rest of simulations presented in this paper, we use the same set of parameter values unless explicitly stated otherwise.

\begin{center}
\begin{table}[t]%
\centering
\caption{Parameter values used in the simulations.\label{tab1}}%
\begin{tabular*}{500pt}{@{\extracolsep\fill}lcccc@{\extracolsep\fill}}
\toprule
\textbf{Parameter name} & \textbf{Value or expression}  & \textbf{Description} \\
\midrule
$k_{m,yc}$ & $2.0$  & Young cancer cell mitosis rate constant   \\
$k_{m,mc}$ & $2.0$  & Mature cancer cell mitosis rate constant   \\
$k_{d,h}$ & $0.15$  & Healthy cell death rate constant   \\
$k_{d,yc}$ & $0.075$  & Young cancer cell death rate constant   \\
$k_{d,mc}$ & $0.075$  & Mature cancer cell death rate constant   \\
$k_{mat}^{cell}$ & $0.05 \cdot k_{m,yc}$  & Cancer cell maturation rate constant   \\
$c_{p}$ & $0.25$  & Threshold nutrient concentration for mitosis   \\
$c_{c_1}$ & $0.2$  & Nutrient concentration regulating cellular death rate  \\
$c_{c_2}$ & $0.1$  & Threshold nutrient concentration for cellular death rate   \\
$k_{occ}^{mv}$ & $0.095$  & Mature vessel occlusion rate constant   \\
$k_{occ}^{yv}$ & $0.135$  & Young vessel occlusion rate constant   \\
$k_{ang}$ & $0.0029449$  & Angiogenesis rate constant   \\
$c_{a}$ & $0.05$  & Threshold nutrient concentration for angiogenesis rate   \\
$p_{crit}^{mv}$ & $0.31$  & Critical pressure for mature vessel occlusion   \\
$p_{crit}^{yv}$ & $0.26$  & Critical pressure for young vessel occlusion   \\
$\epsilon$ & $0.01$  & Threshold ECM volume fraction for angiogenesis rate   \\
$h_{sm}$ & $0.2$  & Smoothness parameter for vessel occlusion function   \\
$D_{c}$ & $1.0$  & Nutrient diffusion coefficient   \\
$k_{c,h}$ & $0.01$  & Consumption rate constant for sustenance of healthy cells   \\
$k_{c,yc}$ & $0.01$  & Consumption rate constant for sustenance of young cancer cells   \\
$k_{c,mc}$ & $0.01$  & Consumption rate constant for sustenance of mature cancer cells   \\
$k_{cm,h}$ & $0.1$  & Consumption rate constant of nutrient used for mitosis of healthy cells   \\
$k_{cm,yc}$ & $k_{cm,h} \cdot k_{m,yc}$  & Consumption rate constant of nutrient used for mitosis of young cancer cells   \\
$k_{cm,mc}$ & $k_{cm,h} \cdot k_{m,mc}$  & Consumption rate constant of nutrient used for mitosis of mature cancer cells   \\
$\Lambda$ & $0.1$  & Tension constant   \\
$\mu_i$ & $10.0$  & Dynamic viscosity   \\
$d_{i,j}$ & $1.0$  & Drag coefficient   \\
$R_{tissue}$ & $16$  & Radius of modelled tissue (circular domain, $\Omega$)   \\
\bottomrule
\end{tabular*}

\end{table}
\end{center}

We impose Dirichlet boundary conditions for Equations~(\ref{eq:massbalancedim_healthy})-\ref{eq:massbalancedim_mv}:

\begin{equation}\label{eq:bcmasssim}
  \theta_i = \theta_i^{\infty},
\end{equation}

\noindent where $\theta_i^\infty$ is the initial volume fraction of phase $i$, i.e.: $\theta_i^\infty = \theta_i(t=0)$.

The boundary condition for nutrient, $c$, is:

\begin{equation}\label{eq:nutrientsim}
\nabla c \cdot \vec{n} = 0.
\end{equation}

\begin{figure}[ht]
\centering
\includegraphics[width = 0.95\textwidth]{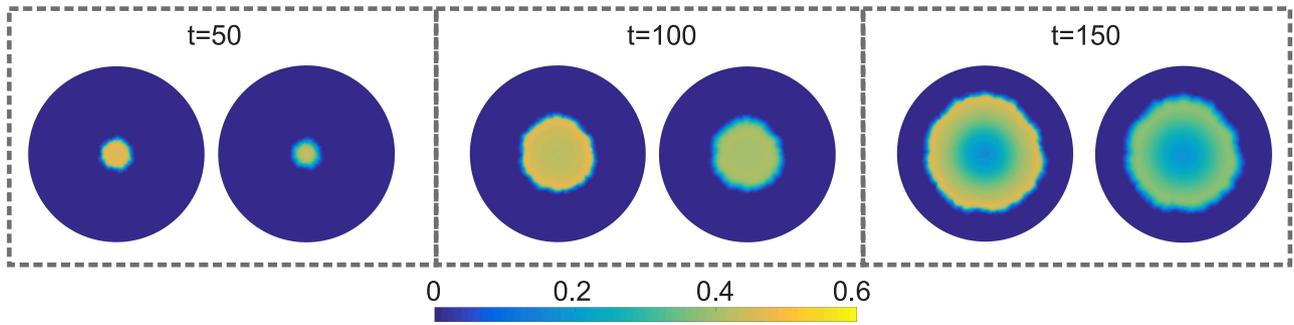}
\caption{Spatial volume fraction distribution of young tumor cells (left panel), and mature tumor cells (right panel) at  $t=50,100$ and $150$. \label{fig:figurematureyoung}}
\end{figure}

Finally, the boundary conditions imposed for the momentum balance Equations~(\ref{eq:mombalancegendim}) are:

\begin{equation}\label{eq:bcmomentumsim}
  \boldsymbol{\sigma}_i \cdot \vec{n} = \vec{0}, i=h,yc,mc,yv,mv
\end{equation}

and Dirichlet (no-slip) condition for ECM:

\begin{equation}\label{eq:bcomomentumsim}
\vec{u}_{ecm} = \vec{0}.
\end{equation}
In this simulation, both cancer cell sub-populations feature double mitosis rate constant, and half death rate constant compared to healthy cells.
As shown in Figure \ref{fig:figurematureyoung}, the population of young cancer cells rapidly starts to expand outwards, while in the same time the maturation process converts part of the young sub-population into mature cancer cells.
The increased proliferation rate of cancer cells (compared to healthy cells) leads to local increases of the cancerous phase density, overcoming natural levels and enabling interactions between cells;
these interactions are the driving force for tumor growth.

\begin{figure}[ht]
\centering
\includegraphics[width=0.63\textwidth]{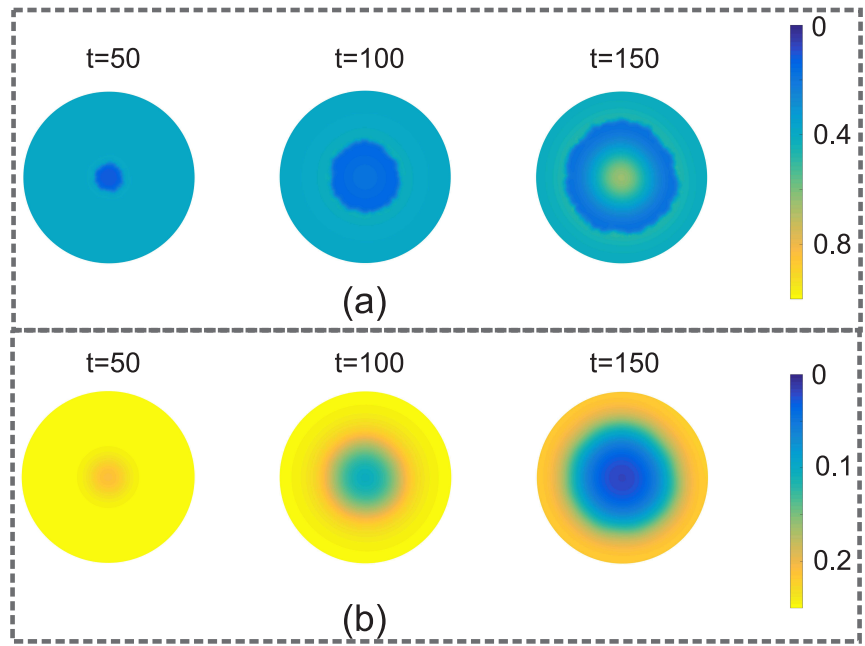}
\caption{(a) Spatial volume fraction of ECM and (b) spatial concentration distribution of nutrient, $c$ at $t=50,100$ and $150$.\label{fig:figECMnutrient}}
\end{figure}
\begin{figure}[ht]
\centering
\includegraphics[width=0.6 \textwidth ]{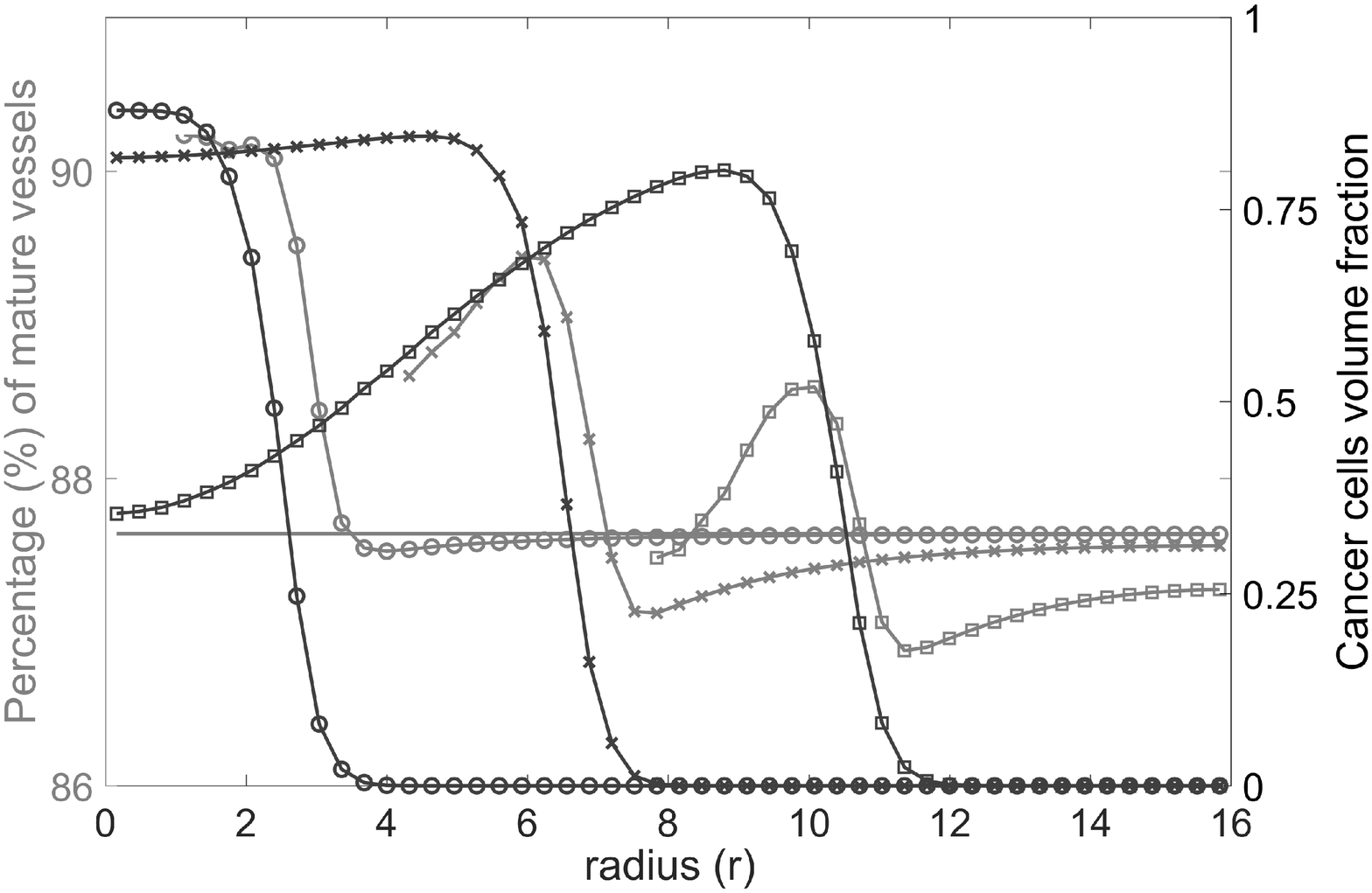}
\caption{Average percentage of mature vessels (grey lines) at  $t=0$, $t=50$ (circled grey line), $t=100$ (crossed grey line) and $t=150$ (squared grey line) along the radial direction of the tissue domain.
Black curves depict the average radial distribution of total cancer cells volume fraction at $t=50,100,150$, (circled, crossed and squared black line, respectively).\label{fig:maturevessels}}
\end{figure}

Furthermore, the relentless proliferation of cancer cells starves the inner region of the tumor from both nutrient and ECM as illustrated in Figure \ref{fig:figECMnutrient}.
However, one can observe the existence of large volume fraction values of ECM at $t=150$.
This is attributed to the formation of a necrotic zone in the interior of the tumor.
In an environment lacking the nutrients essential to maintain live cells, the cells in the interior of the tumor succumb to necrosis forming the tumor’s necrotic zone.
The increased cancer cell volume fractions lead to increases in the exerted pressure around the tumor.
As a result, the vasculature in the vicinity of the tumor breaks down, thus further aggravating the problem of nutrient transport to the tumor's interior.
As reported previously, mature vessels feature the ability to withstand increased pressure values compared to young sprouts.
This increase in the resilience of mature vessels is illustrated in Figure \ref{fig:maturevessels}.
In particular, we depict snapshots of the average ratio of mature vessels over the total volume fraction of vessels along the radial direction (left vertical axis), complemented by the average radial distribution of cancer cells volume fractions (right vertical axis).
One can observe that at the vicinity of the tumor's front the vessels are predominately mature, as a result of their increased resilience.
Furthermore, throughout the exterior region of the tumor, the balance shifts in favor of the younger vessels, due to the angiogenesis process.

\begin{figure}[ht]
\centering
\includegraphics[width=0.6 \textwidth ]{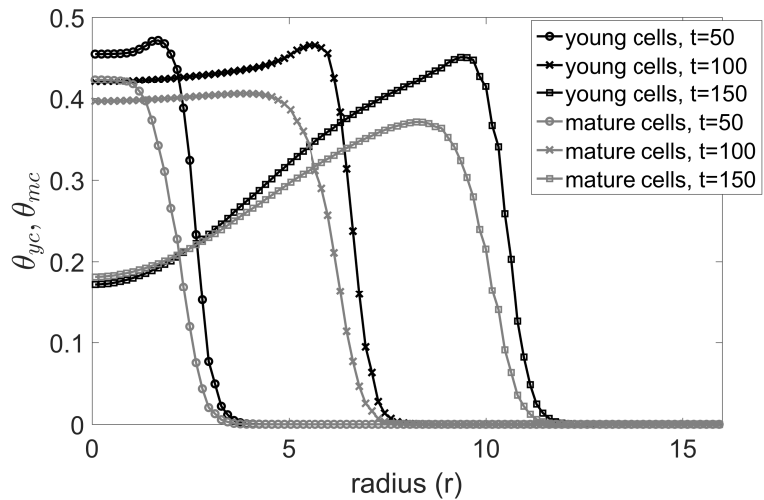}
\caption{Snapshots of the mean radial distribution of young cancer cells (black lines) at  $t=50$ (circled black line), $t=100$ (crossed black line) and $t=150$ (squared black line) along the radial direction of the circular domain, $\Omega$.
Grey lines depict the mean radial distribution of mature cancer cells volume fraction at $t=50,100,150$, (circled, crossed and squared grey line, respectively).\label{fig:radialmatureyoung}}
\end{figure}

Figure \ref{fig:radialmatureyoung} depicts the average radial distribution of young and mature cancer cells at different time instances.
Interestingly enough, younger cells form an outer layer acting as a protective shield for mature cells.
Figure \ref{fig:radialmatureyoung} further establishes that the different age group cells are unevenly distributed along the radial direction of the developed tumor.
Mature cells mostly inhabit the quiescent and necrotic zones of the tumor, while the proliferating zone’s cancer cells are predominantly young.

\begin{figure}[ht]
\centering
\includegraphics[width=0.92\textwidth]{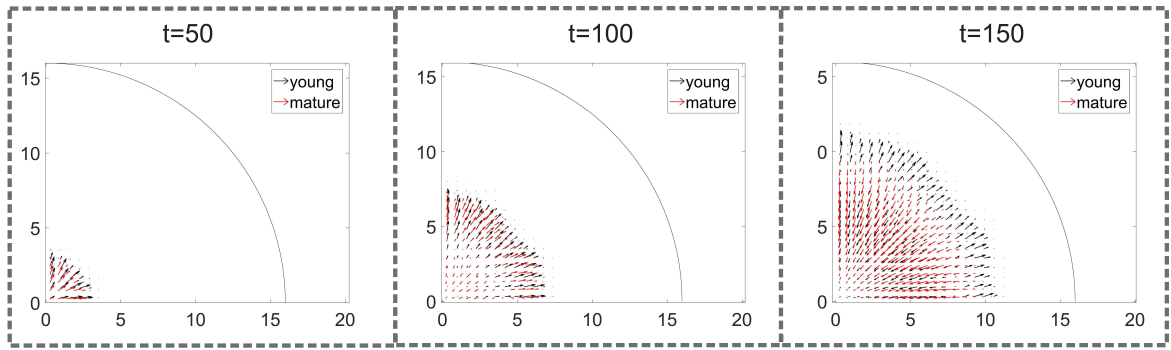}
\caption{Evolution of cancer cell phases fluxes,$\theta_i \cdot \vec{u}_i, i=yc,mc$ for $t=50,100$ and $150$.
The vector length is proportional to its magnitude.
\label{fig:figureflux}}
\end{figure}

Due to the uncontrolled proliferation rate of cancer cells, one can observe the formation of a macroscopic velocity field.
In general, they initially present an outward-pointing motion in an effort to relieve the built up pressure.
After the formation of the quiescent and the necrotic core, cancer cells start to perform a second motion.
Due to the volume fraction gradient and the consequent pressure gradient developed between the proliferation zone and the rest of the tumor zones, cancer cells in the interior of the tumor rush inwards to smoothen that gradient and inevitably succumb to necrosis.
The flux $\theta_i \cdot \vec{u}_i$ of young and mature cancer cells ($i=yc,mc$, respectively) is illustrated in Figure \ref{fig:figureflux} at different time instances.
During the initial stages, both phases exhibit an outward-pointing motion;
young cancer cells preserve this behavior even at later stages, whereas mature cancer cells gradually present the tendency of moving towards the tumor’s inner core as time progresses.
It is thus evident that the absence of uniformity in spatial distribution of young and mature cancer cells originates from the different motility of the two sub-populations.

\subsection{Effect of mitosis rate heterogeneity}\label{sec:mitosis}
\begin{figure}[ht]
\centerline{\includegraphics[width=0.6\linewidth]{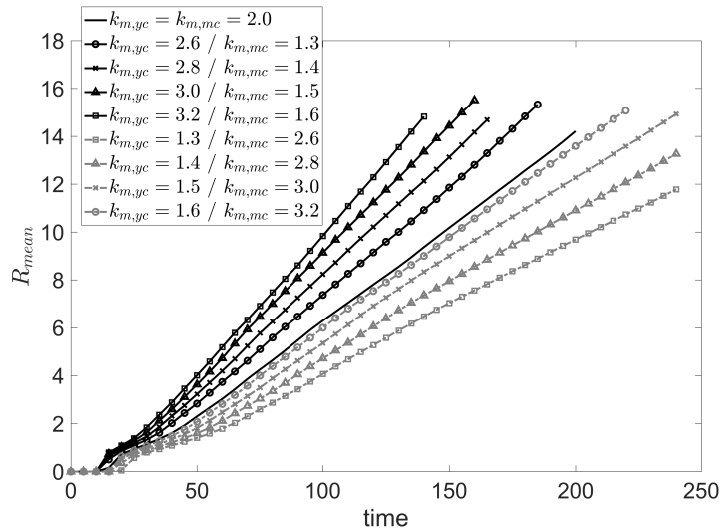}}
\caption{Evolution of $R_{mean}$ for various combinations of $k_{m,yc}-k_{m,mc}$ values.
\label{fig:comparison_mitotic}}
\end{figure}

A measure to assess the growth rate of a tumor is by computing the average distance of the tumor's front from the origin of the domain, $R_{mean}$.
In particular, we determine $R_{mean}$ as the average distance of contour, $\theta_{yc}+\theta_{mc}=\theta^*=0.6$ from the origin of the domain.
In Figure~\ref{fig:comparison_mitotic}, we depict the evolution of $R_{mean}$ for different mitosis rate constants of young and mature cancer cells.
Tumors with young cancer cells featuring higher rates of mitotic activity, appear to grow faster.

This behavior comes in agreement with Figure \ref{fig:figureflux} showing that young cells are primarily responsible for the tumor’s expansion.

\begin{figure}[ht]
\centerline{\includegraphics[width=0.6\linewidth]{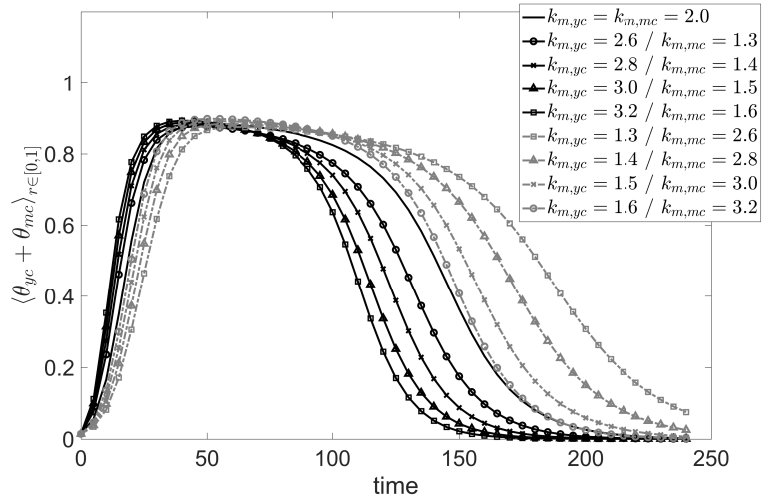}}
\caption{Evolution of average cancer cells volume fraction in the area of the initial cancerous seed ($\sqrt{x^2+y^2} \leq R_o=1$) for different combinations of $k_{m,yc}$ and $k_{m,mc}$ values.\label{fig:canc_volfrac_mitosis}}
\end{figure}
\begin{figure}[ht]
\centering
{\includegraphics[width=0.52\linewidth]{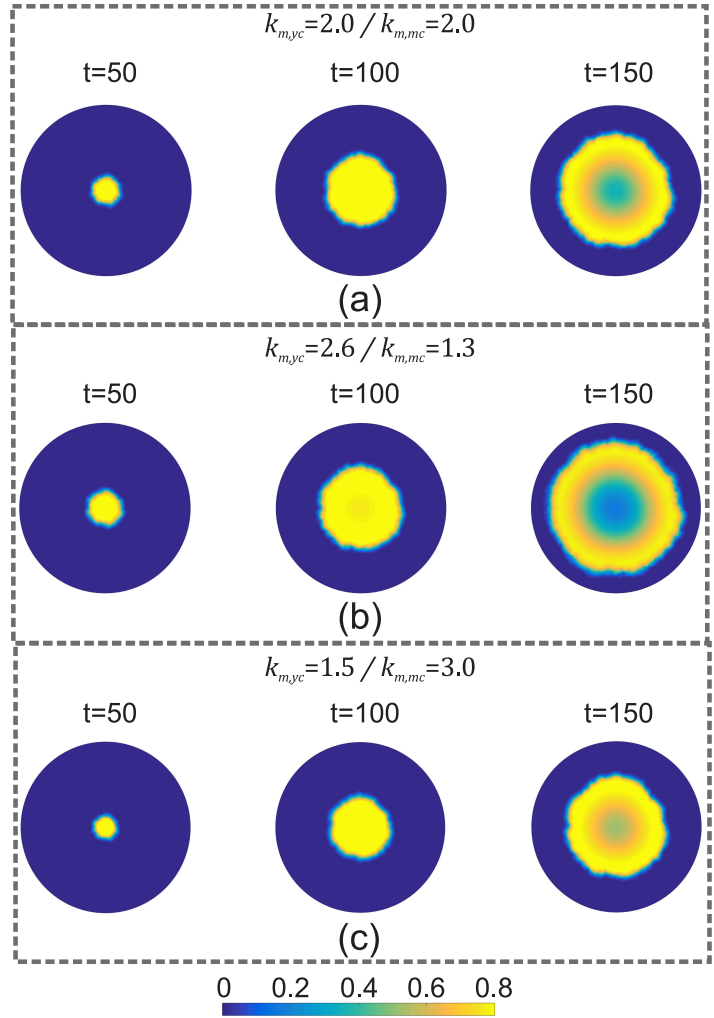}}
\caption{Spatial volume fraction distribution for tumor cells ($\theta_{yc}+\theta_{mc}$) with (a) $k_{m,yc}=k_{m,mc}=2.0$ (top), (b) $k_{m,yc}=2.6 - k_{m,mc}=1.3$ (middle) and  (c) $k_{m,yc}=1.5 - k_{m,mc}=3.0$ (bottom) for $t=50,100$ and $150$.
\label{fig:canc_mit_rates}}
\end{figure}

It is then to be expected from what is shown in Figure \ref{fig:figureflux} that tumors with more active mature cells will show a delay in the formation of the necrotic core compared to the rest.
Indeed, as shown in Figure \ref{fig:canc_volfrac_mitosis}, these tumors exhibit a significant delay in the necrotic core formation.
Furthermore, Figure \ref{fig:canc_volfrac_mitosis} illustrates the evolution of cancer cells volume fraction in the area of the initial seed ($r = \sqrt{x^2+y^2} \leq 1$).
Cells in the tumor’s interior evade necrosis for longer periods of time, specifically in cases where the overall mitosis rate has relatively lower value and cancer cells do not starve their environment from nutrients at the same rate.
As expected from the above observations, the fastest growing tumor is the one with young cells featuring more intense mitotic activity.
A more detailed view of the tumor’s morphology is presented in Figure \ref{fig:canc_mit_rates} for three representative scenarios: (a) equal mitosis rate constant values for young and mature cells, (b) higher mitosis rate constant for younger cells, and (c) lower mitosis rate constant for younger cells.
The tumor with $k_{m,yc}=2.6$ and $k_{m,mc}=1.3$ grows faster compared to the case: $k_{m,yc}=1.5 - k_{m,mc}=3.0$, and develops earlier a necrotic core.

\begin{figure}[ht]
\centerline{\includegraphics[width=0.6\linewidth]{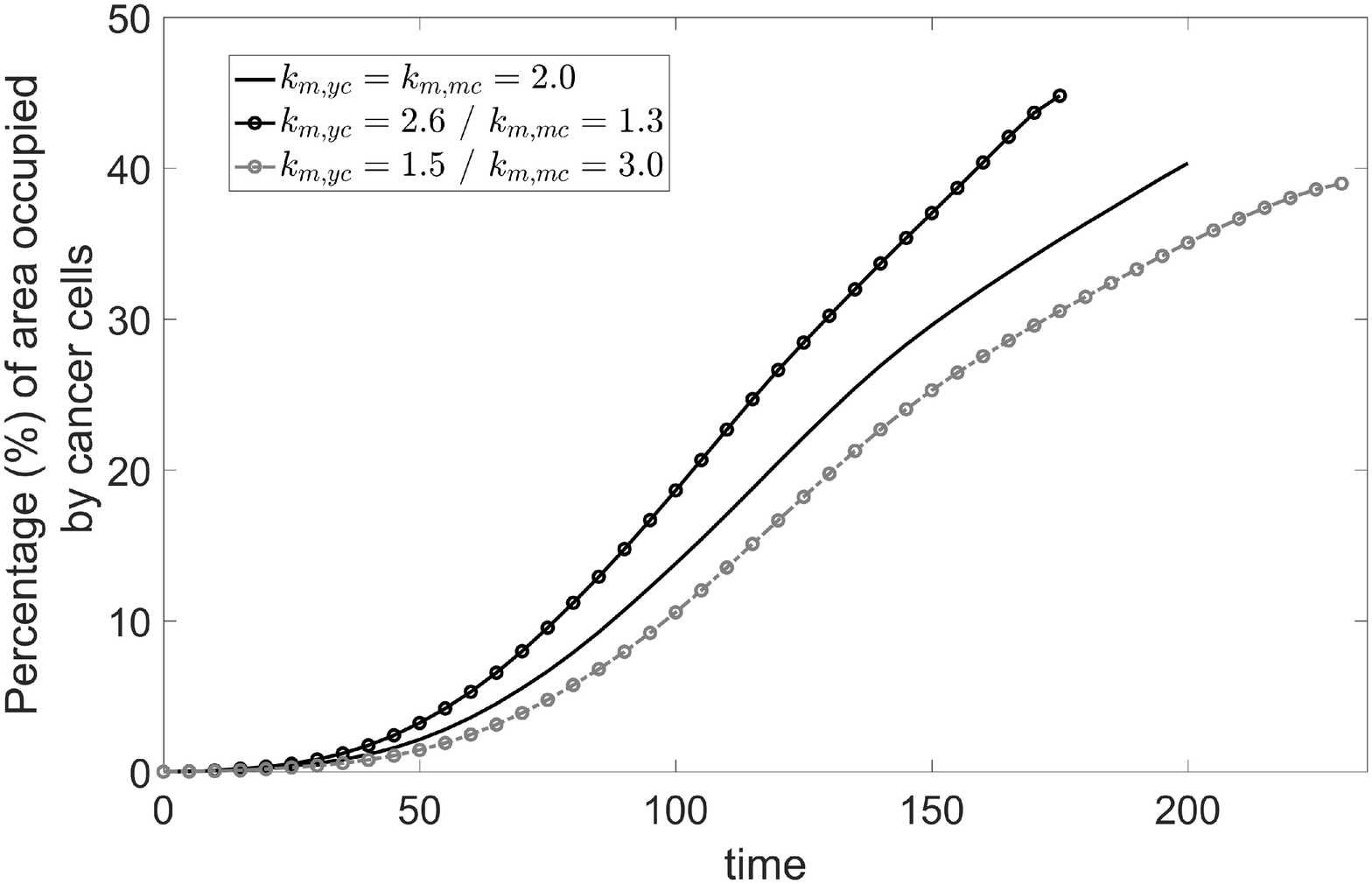}}
\caption{Temporal evolution of the domain’s area percentage covered by cancer cells for tumor with equivalently proliferating cancer cells sub-populations ($k_{m,yc}=k_{m,mc}=2.0$), tumor with more proliferative young cells ($k_{m,yc}>k_{m,mc}$), and tumor with more proliferative mature cancer cells (($k_{m,yc}<k_{m,mc}$). \label{fig:perc_coverage_mitosis}}
\end{figure}

Finally, we quantify the aggressiveness of the growing tumors for different combination values of mitosis rate constants, by computing the total surface covered by cancer cells at different time instances.
In particular, we compute the surface integral of the cancer cell volume fractions on the computational domain over the surface of the said domain, $S = \pi R_{tissue}^2$:

\begin{equation}\label{volumefraction}
  a \% = \frac{100 \%}{\pi R_{tissue}^2} \iint_S \left( \theta_{yc} + \theta_{mc} \right) \mathrm{d}x\mathrm{d}y.
\end{equation}
Figure \ref{fig:perc_coverage_mitosis} shows the evolution of $a \%$ for three representative cases of mitosis constant rate values for young and mature cancer cells, and reveals the higher aggressiveness of tumors with young cells exhibiting the highest mitosis rate.

\subsection{Effect of cellular death rate heterogeneity}
\begin{figure}[ht]
\centerline{\includegraphics[width=0.6\linewidth]{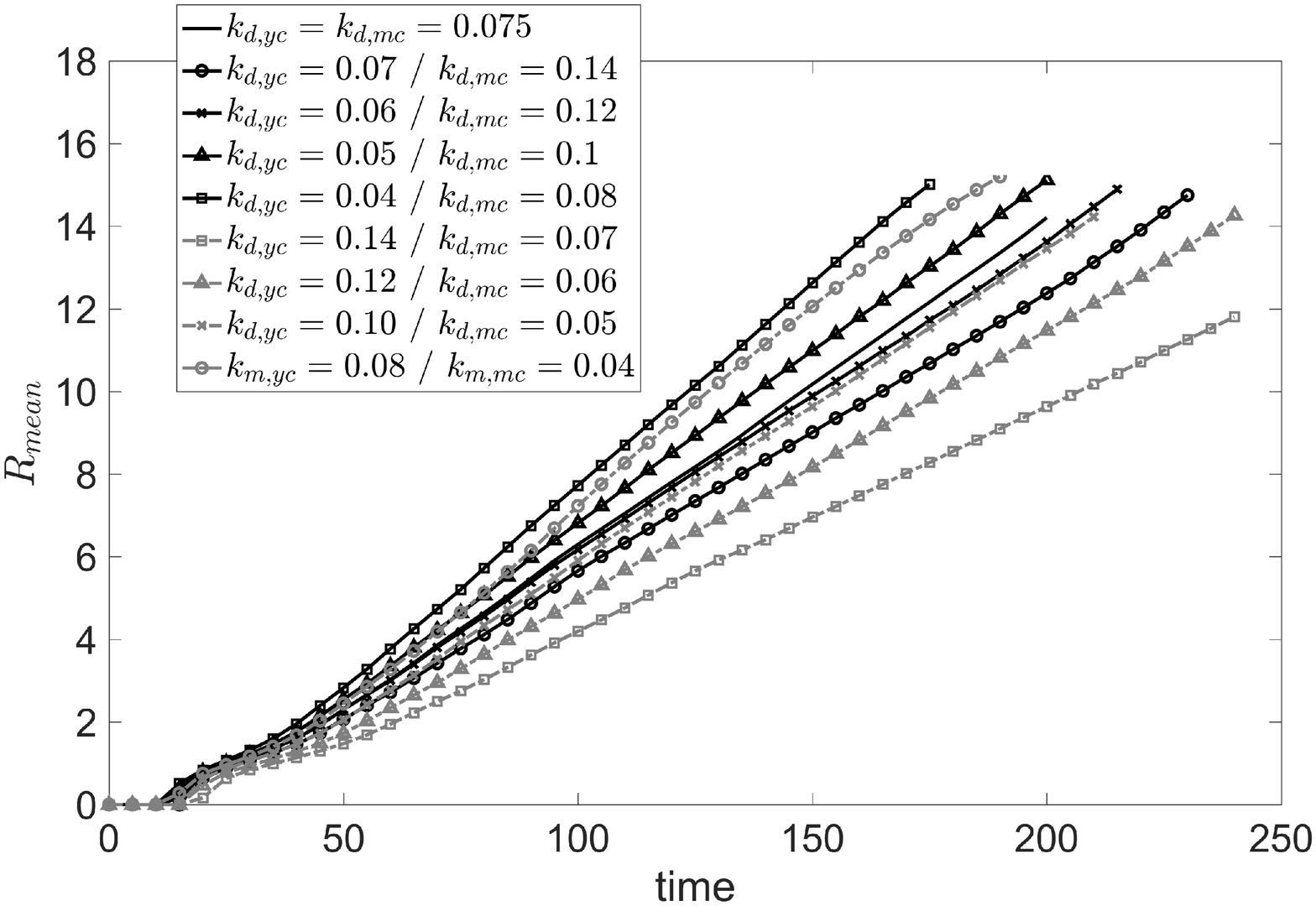}}
\caption{Evolution of $R_{mean}$ for various combinations of $k_{d,yc}-k_{d,mc}$ values.
\label{fig:canc_volfrac_necrosis}}
\end{figure}

In the present paragraph, we present the effect of the death rate constant of young and mature cancer sub-populations.
Here, the ratio of death rate constants $k_{d,yc}$ and $k_{d,mc}$, for young and mature cancer cells, respectively, is kept constant.
$\frac{k_{d,yc}}{k_{d,mc}} =2$ for tumors where young cells are more susceptible to necrosis and $\frac{k_{d,yc}}{k_{d,mc}} = \frac{1}{2}$ for tumors with young cancer cells being more resilient compared to mature cells.
$R_{mean}$ is calculated for a series of simulations covering a range of $k_{d,yc}$ and $k_{d,mc}$ combinations satisfying both ratios.
The results of these calculations is presented in Figure \ref{fig:canc_volfrac_necrosis}.
Tumors with more resilient young cells show a tendency to expand at a higher rate, however one can observe that for the studied cases the differences are not as apparent as in the results presented in the previous paragraph, where the effect of mitosis rate heterogeneity is examined.

\begin{figure}[ht]
\centerline{\includegraphics[width=0.52\linewidth]{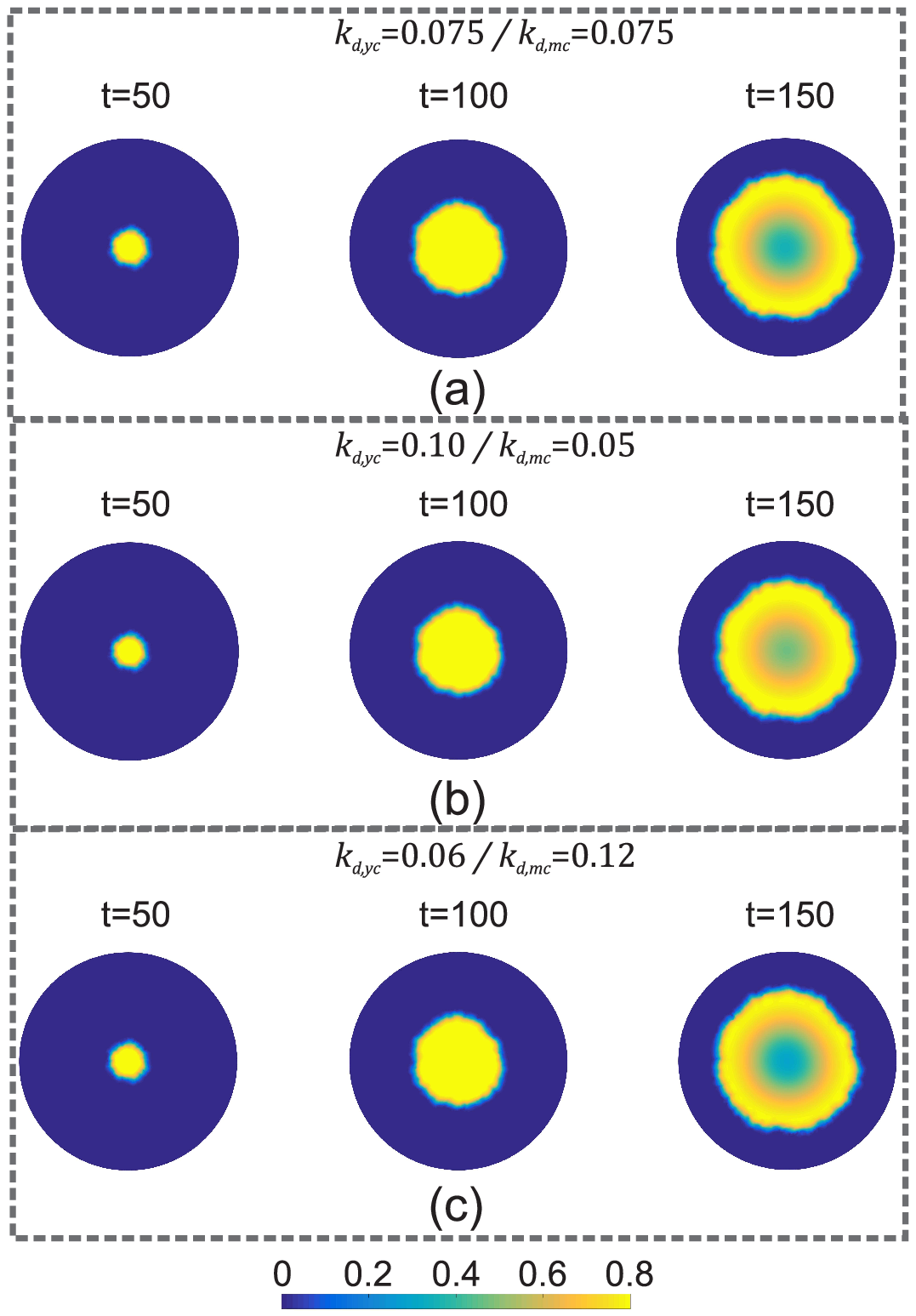}}
\caption{Spatial volume fraction distribution for tumor cells with (a) $k_{d,yc}=k_{d,mc}=0.075$ (top), (b) $k_{d,yc}=0.1 - k_{d,mc}=0.05$ (middle) and  $k_{d,yc}=0.06 - k_{d,mc}=0.12$ (bottom) for $t=50,100$ and $150$.
\label{fig:canc_death_rates}}
\end{figure}

Despite the fact that the growth rates do not show significant differences for the different combinations of death rate constants, the examination of the resulting tumor morphology shows interesting dissimilarities.
Figure \ref{fig:canc_death_rates} illustrates snapshots of the cancer cell volume fraction for different death rates of young and mature cancer cells.
All three tumors grow at a similar rate;
however, significant differences in terms of the formation of a necrotic core can be observed.
In particular, tumors with more resilient mature cancer cells tend to form a necrotic core of smaller size compared to the rest of the studied cases at $t=150$.
This comes in agreement with the observation that mature cells are primarily focused in replenishing the dead cells in the inner region of the tumor.

\begin{figure}[ht]
\centerline{\includegraphics[width=0.6\linewidth]{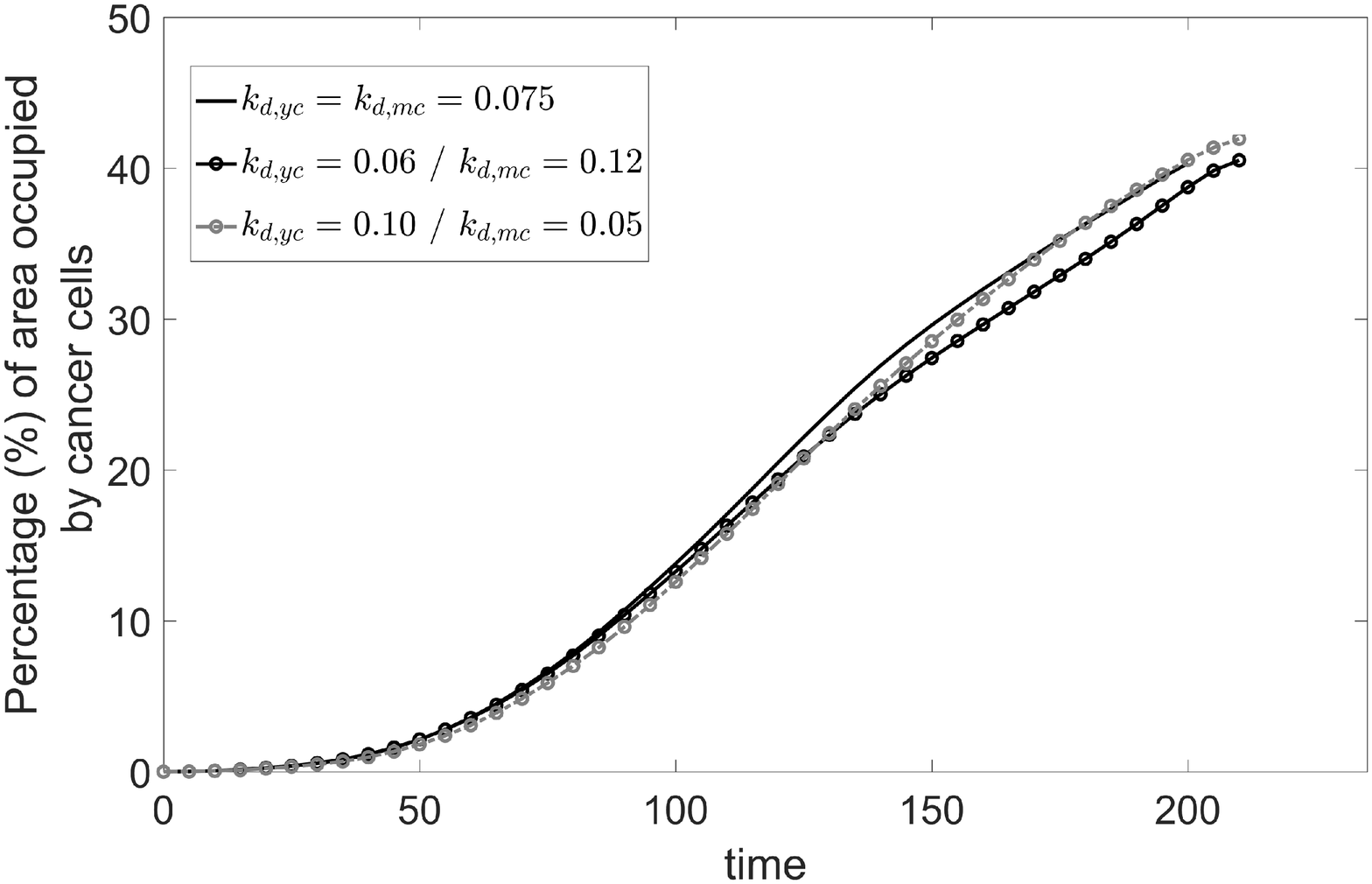}}
\caption{Temporal evolution of the domain’s area percentage, $a \%$, covered by cancer cells for heterogeneous tumors with $k_{d,yc}=k_{d,mc}=0.075$, $k_{d,yc}=0.1 - k_{d,mc}=0.05$ and  $k_{d,yc}=0.06 - k_{d,mc}=0.12$. \label{fig:perc_coverage_necrosis}}
\end{figure}

Finally, we quantify the aggressiveness of the developing tumor, by comparing the fraction of surface that is covered by cancer cells, $a \%$, using Equation~(\ref{volumefraction}).
The results are presented in Figure \ref{fig:perc_coverage_necrosis} for three representative scenarios: (a) equal death rate constants, (b) more resilient mature cells, and (c) more resilient young cells confirming our previous ascertainment that the growth rates of the heterogeneous tumors are similar when varying the death rate constants of young and mature cancer cells.

\subsection{Effect of material properties}
\begin{figure}[ht]
\centerline{\includegraphics[width=0.95\linewidth]{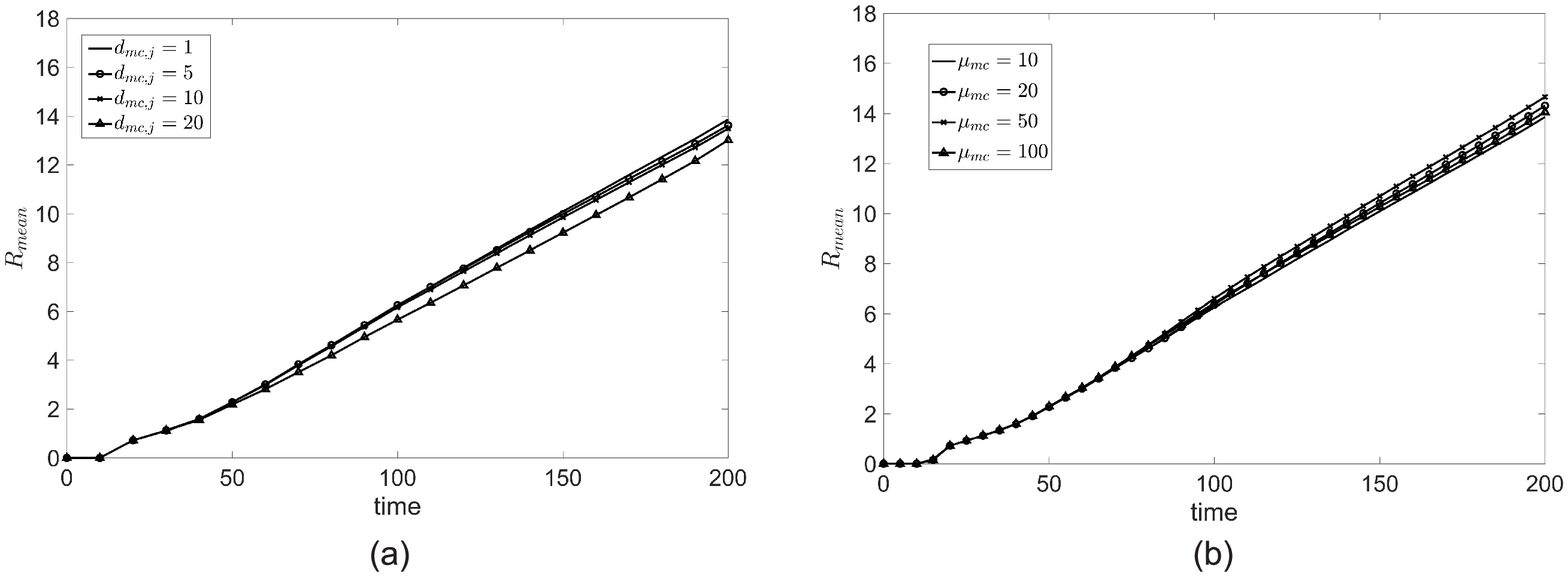}}
\caption{Evolution of $R_{mean}$ for various values of mature cancer cells (a) drag coefficients $d_{mc,j}$ (left) and (b) dynamic viscosity $\mu_{mc}$ (right). \label{fig:comparison_material}}
\end{figure}

In this paragraph, we assess the tumor growth dependency on the material parameters of cancerous fluid phase;
in particular, we examine the effect of the viscosity and drag coefficients values for the mature cancer cell sub-population.
This assessment is performed by computing $R_{mean}$, which is defined in paragraph~\ref{sec:mitosis}.
Figure \ref{fig:comparison_material}(a) depicts the growth rate of tumors for different values of the drag coefficients, $d_{mc,j}$, in the interphase between mature cancer cells and the rest of the fluid phases ($j=h,yc,yv,mv,ecm$).
By increasing the interphase drag results in a delay in the tumor growth;
such an effect is to be expected since increases in $d_{mc,j}$ bolster the force, which opposes to the cells expansion movement.
Figure \ref{fig:comparison_material}(b) shows the evolution of $R_{mean}$ for various values of the dynamic viscosity $\mu_{mc}$ of mature cancer cells.
Here, the trend of tumor growth rate with changes in $\mu_{mc}$ is not as clear as in the case of interphase drag.
When the mature cancer fluid phase is behaving as a highly viscous one, an increase in viscosity does not yield a clear trend in the obtained tumor dynamics.
The highest growth rate is attained for $\mu_{mc}=50$, whereas the lowest growth rate for $\mu_{mc}=10$.
The simulation of the highest studied viscosity value $\mu_{mc}=200$ results the second lowest tumor growth rate (exceeds only the $\mu_{mc}=10$ case).
In addition, by observing the equal growth rates at the initial stages of the simulations for different interphase drag coefficients and viscosity values, one can infer that the time needed by the tumor to surpass natural cell density and start expanding is not affected by the mature cells material properties.

\begin{figure}[ht]
\centerline{\includegraphics[width=0.95\linewidth]{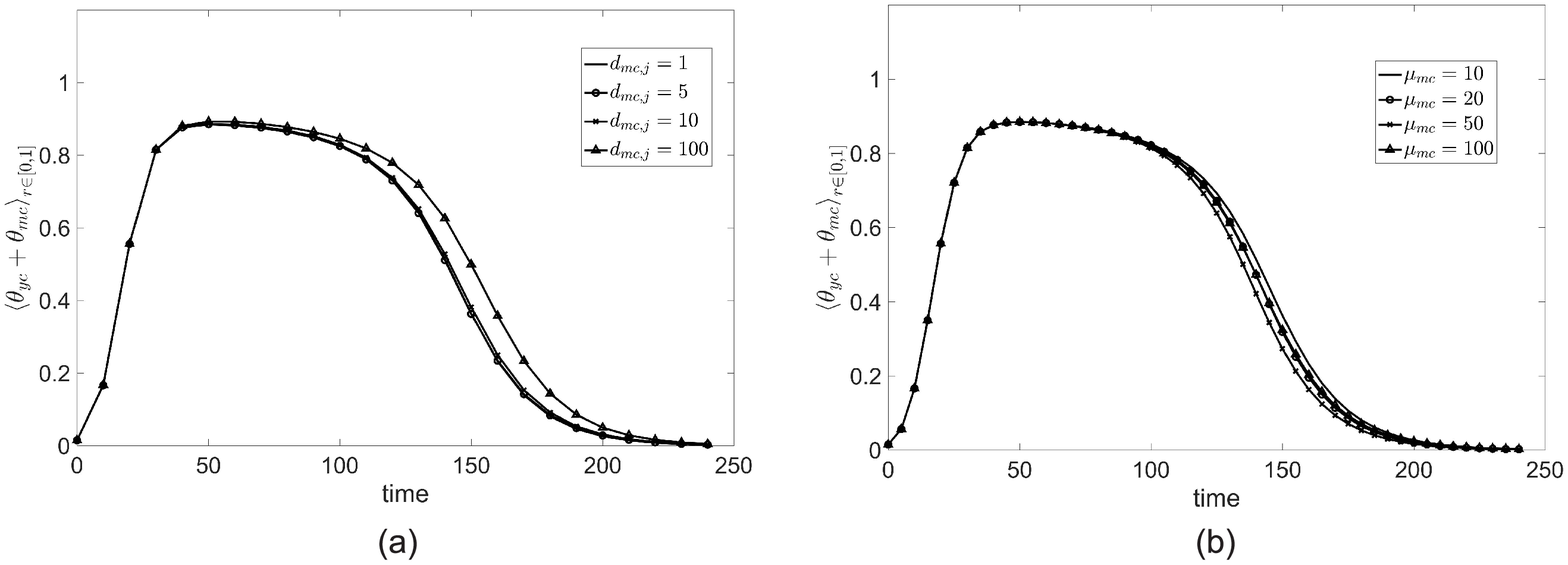}}
\caption{Average cancer cells volume fraction in the area of the initial cancerous seed ($r=\sqrt{x^2+y^2} \leq 1$) for different values of $d_{mc,j}$ (left) and $\mu_{mc}$ (right).\label{fig:canc_volfrac_material}}
\end{figure}

We also study the effect of the cellular material properties of mature cancerous sub-population on the tumor's morphology, and in particular the size of the developed necrotic core.
Figure \ref{fig:canc_volfrac_material} presents the average volume fraction of cancer cells within a radius $r=\sqrt{x^2+y^2} \leq 1$ from the origin of the tissue domain.
For values of $d_{mc,j}$, ranging from $1$ to $10$, there is no clear effect on the time of formation of a necrotic region.
By further increasing its value to $d_{mc,j}=100$ one can observe a delay in the formation of the necrotic core.
Again, an increase of the viscosity of mature cancer cells fluid phase does not seem to alter significantly the time of formation,  as well as the size of the necrotic core.

\section{Discussion and Conclusions}\label{discussion}

In this work, we present a continuum-level computational model, which simulates the complex and unstable environment of a growing heterogeneous malignant tumor by treating the host tissue as a multi-phase fluid mixture.
The model features flexibility, since it can be easily modified to incorporate additional elements of heterogeneity, with relatively low computational requirements.
In particular, the present study adopts the modeling concept presented in \cite{Hubbard:2013}, and incorporates the effect of ITH by introducing sub-populations of cancer cells, which represent different stages of cellular maturity.
In addition, the microvasculature is categorized into mature blood vessels and young sprouts equipped with different levels of ability to withstand vascular occlusion.
The resulting system of partial differential equations formulating mass and momentum balances for the different phases of the mixture is solved with the Finite Elements Method utilizing the commercial software Comsol Multiphysics \textsuperscript \textregistered.
By observing the dynamics of the different sub-populations involved in the model, the young cancer cells are mostly responsible for the expansion of the tumor;
during the course of simulations the young cells form a protective layer shielding the inner layers of the tumor, where mature cells are more abundant.
Meanwhile, mature cells move towards the interior of the tumor due to the pressure exerted by the cells in the proliferating rim.
The concept of ITH was further explored by testing different combination values of kinetic parameters so as to determine whether a tumor with heterogeneous properties proliferates at a different rate compared to a tumor whose kinetic parameters are uniform throughout its cancer cell sub-populations.
Indeed, our simulations reveal that introducing heterogeneity in the cellular mitosis rate holds a crucial role in the resulting
dynamics of the tumor in question.
When young cancer cells exhibit higher mitotic activity compared to mature cells, the overall dynamics of the growing tumor are considerably faster as opposed to cases where the mature cells are considered to proliferate at higher rates.
This observation can be considered in congruence with the approach of differentiation therapy for malignant tumors, a therapy based on
forcefully differentiating cancer cells into a specific mature state \cite{Wang:2013,Leszczyniecka:2001}.
We adopt a similar approach in order to associate ITH with cellular death rates.
In this case, we observe again changes in the macroscopic tumor dynamics;
however, the computed growth rates for different combinations of cellular death rates are not as significant as in the case of modifying their proliferation rate.
We believe that our study is amenable on further modifications in order to assess various topics related to ITH, and its implications to therapeutic strategies.
A possible enhancement to our model can be made by substituting the existing kinetics in order to model the cellular metabolism to incorporate a larger array of
phenomena, such as the interaction of the cells with different chemicals and nutrients or separating apoptosis from necrosis.
In addition, our model can be extended to incorporate possible treatments and therapeutic strategies, including chemotherapy, radiotherapy, virotherapy or hyperthermic techniques.
Apart from evaluating the efficacy of these techniques individually, their synergy with certain chemical agents or elements of the tumor’s micro-environment can
also be simulated (e.g., interaction with macrophages).
Our modeling approach to sub-populate cancer cells depending on their state of maturity, can be trivially modified in order to incorporate the effect of differentiation therapies \cite{Wang:2013,Leszczyniecka:2001}, with the inclusion of appropriate chemical agents that induce the maturation process.
Furthermore, the inclusion of cancer stem cells (CSC) as being one of the cancerous phases during the study of a therapeutic path is an exciting new
prospect since,  CSC are often responsible for provoking cancer relapse due to their –at least partial- immunity to certain therapies \cite{Sontag:2020,Clevers:2011}.
Another possible extension of the present model is to incorporate the effect of macrophages, i.e., the immune cells most common in the immediate tumor environment \cite{Unver:2019}.
Cancer cells have the ability to secrete chemical compounds that mediate macrophage chemotaxis,
i.e., direct the movement of macrophages based on these chemicals’ concentration gradient.
%





\subsection*{Conflict of interest}

The authors declare no potential conflict of interests.

\bibliography{wileyNJD-AMA}%

\clearpage

\end{document}